\documentclass[superscriptaddress,11pt]{revtex4}
\usepackage{doi}
\usepackage{hyperref}
\hypersetup{
  colorlinks=true,        
  linkcolor=blue,         
  citecolor=cyan,         
}

\usepackage{graphicx}
\usepackage{dcolumn}
\usepackage{bm}
\usepackage{color}
\usepackage{enumitem}
\usepackage{amsmath}
\usepackage{amssymb}

\begin{document}

\title{Particle Dynamics and Quasi-Periodic Oscillations in the GUP-Modified Schwarzschild Spacetime: Constraint Using Micro-Quasars Data}

\author{Husanboy Hoshimov}

\affiliation{Fergana State University, Murabbiylar St 19, Fergana 150100, Uzbekistan}
\affiliation{National Research University TIIAME, Kori Niyoziy 39, Tashkent 100000, Uzbekistan}
\affiliation{Fergana Polytechnic Institute, Fergana street 86, Fergana 150107, Uzbekistan}

\author{Odil Yunusov}

\affiliation{Inha University in Tashkent, Ziyolilar 9, Tashkent 100170, Uzbekistan}

\author{Farruh Atamurotov}

\affiliation{Inha University in Tashkent, Ziyolilar 9, Tashkent 100170, Uzbekistan}
\affiliation{University of Tashkent for Applied Sciences, Str. Gavhar 1, Tashkent 100149, Uzbekistan}
\affiliation{Urgench State University, Kh. Alimdjan str. 14, Urgench 220100, Uzbekistan}

\author{Mubasher Jamil}
\email{mjamil@sns.nust.edu.pk (corresponding author)}
\affiliation{School of Natural Sciences, National University of Sciences and Technology, Islamabad, 44000, Pakistan}

\author{Ahmadjon Abdujabbarov}
\affiliation{National Research University TIIAME, Kori Niyoziy 39, Tashkent 100000, Uzbekistan}
\affiliation{Shahrisabz State Pedagogical Institute, Shahrisabz Str. 10, Shahrisabz 181301, Uzbekistan}
\affiliation{Tashkent State Technical University, Tashkent 100095, Uzbekistan}

\date{\today}

\begin{abstract}
In this work, we have worked out dynamical aspects for the particles moving around the GUP-corrected-Schwarzschild (S-GUP) black hole. We have calculated the innermost stable circular orbit (ISCO) around black hole and explored its implications for different microquasars. Additionally, we have shown that the Kerr black hole mimics S-GUP black hole after some tuning of parameters. Finally, considering the S-GUP black hole as a microquasar source, we have studied quasi-periodic oscillation (QPO). Further utilizing the available observational data of few microquasars, we have obtained constrains on the GUP parameter $\epsilon$ as well. 
\end{abstract}
\maketitle
\newpage
\section{Introduction} 

In theoretical physics, the problem of unifying quantum mechanics with gravity remains a most challenging problem. 
One of the numerous approaches to resolve the issue to the quantum gravity, the thermodynamics of a black hole (BH) serves as superior model to achieve the goal of unifying gravity and quantum mechanics. This is due to the fact that a BH is not only gravitational entity but also a quantum mechanical object. They BHs serve as unique laboratory for testing predictions of quantum gravity and offer a theoretical framework for advancing towards the goal of a theory of quantum gravity \cite{Susskind:1995qc}. 

The generalized uncertainty principle (GUP) serves as a valuable tool to address the limitations of general relativity (GR). The uncertainty relationship between position and momentum operators due to GUP incorporates the nonlinear terms due to uncertainties in momentum $([\hat{x},\hat{p}]=\iota \hbar (1+\beta \hat{p}^2))$ which are motivated both from string theory and loop quantum gravity \cite{Ali:2009zq}. The GUP suggests a minimal length scale at the Planck level, which effectively eliminates the singularity predicted by standard General Relativity. Following this discovery, physicists have extensively explored the implications of GUP \cite{Nasser2015RPPh}. GUP has been approached in various ways, including formulations rooted in string theory \cite{Veneziano1986EL,Amati1987PhLB,Amati1989,Gross1987PhLB,Gross1988NuPhB}. In literature, various astrophysical tests of the GUP corrected BH  have been considered including black hole shadows and QPOs. In addition, estimates of the  GUP parameter have been linked to various physical phenomena. These include constraints derived from gravitational wave events, calculations of perihelion precession for planets in the solar system and binary pulsars, analysis of shadows via Shapiro time delay, investigations of gravitational red-shift, and consideration of geodetic precession within the framework of the GUP-modified Schwarzschild metric \cite{Adler2001GReGr,2020EL,Karmakar2022IJMPA,Fu2021NuPhB,Li2017GReGr,Rizwan2017IJMPD,Jusufi:2020wmp}. The tests also include studying the effect of GUP on the accretion disk onto Schwarzschild black hole \cite{Moussa:2023}.   Moreover, research has delved into the study and constraints of GUP-corrected Kerr black holes through QPOs and the motion of the S2 star around the Milky Way galactic center. Additionally, esoteric objects such as wormholes have been explored through the inclusion of GUP corrections in calculations of Casimir energy density \cite{Jusufi:2020rpw}.

In order to justify any modifications or alternatives to GR, analyzing the dynamics of test particles is the most efficient and convenient way to constrict and apply such tests. By analyzing the motion of massive and/or massless particles in any metric theory of gravity, we can gain valuable insights into the physical parameters of the solution~\cite{Bambi2017book,Chand:1998a}. 
It is worth mentioning that X-ray data from astrophysical objects can also be used to test solutions of gravity models~\cite{Wilkins2012MNRAS}. 
Analysis of the motion of massless particles can be used to investigate the phenomenon of gravitational lensing, which is considered the most featured characteristics of GR. Specifically, researchers have examined the effects of gravitational weak lensing around compact objects in the presence of plasma environment in reference~\cite{GUP}.

Studying the special case of particle motion that corresponds to circular orbits can be an effective method for examining the astrophysical phenomenon called QPOs. QPOs are believed to be observed in compact objects that have matter flowing towards them from a companion star, which can be detected in the X-ray band. For the first time, QPO has been detected by analyzing the fluctuations in the power spectra of flux from X-ray binary pulsars~\cite{1989ApJ...346..906A}. 
The detection of QPOs prompted numerous authors to suggest various models to explain them quantitatively and qualitatively. Out of the many models available, the most promising ones rely on the characteristics of circular orbits of test particles. The concept is straightforward: the innermost stable circular orbits (ISCOs) may give rise to fundamental frequencies and, consequently, the frequencies of QPOs. As ISCO is linked to the strong field regime around compact objects, QPOs can be utilized to test gravity models in the strong field regime, resulting in constraints on the free parameters of different gravity models based on the QPO data. 
We refer readers to Refs.~\cite{Wagoner2021RNAAS,Abramowicz2011AcA,
Wang2024PatRe,Ingram2021MNRAS,
Fragile2020MNRAS,Stuchlik2013AA,Ortega-Rodriguez2020MNRAS,Maselli2020ApJ,Rayimbaev2021PhRvD,
2024PhyS...99f5011A,2024JHEAp..43...51D,2024ChPhC..48e5104R,2024JHEAp..43..158F}, for review of models of QPOs based on circular orbits of test particles. Here, we plan to analyze the GUP-modified Schwarzschild solution with the aim to develop a test using QPO observations. 

One of the leading frontiers in modern physics research is the experimental testing of black holes motivated by quantum gravity (QG) effects using observations of various astronomical sources. The aim is always to test and identify the role played by the relevant quantum gravity parameters in the dynamics of light and matter in strong gravitational fields. These phenomenological investigations help in identifying which QG corrections are consistent (or in conflict) with observational data. In literature, various QG motivated are tested using different observational or experimental methods such as X-rays, Iron line spectroscopy, binary BHs and gravitational wave, BH shadows, binary pulsars and several solar system tests \cite{Rayimbaev:2020jye,Zhu:2020cfn,Kumar:2022vfg,Vagnozzi:2022moj}. In this context, GUP corrected BH is also motivated by quantum gravity approaches and it involves a free parameter $\epsilon$. Previously, this black hole 
has been tested with solar system tests \cite{Scardigli2015EPJC,2024PDU....4301392H}, and we are mainly interested in testing this metric with QPOs and providing constraints on the parameter $\epsilon$. 
In this work, we have worked out dynamical aspects for the particle moving in the GUP-corrected-Schwarzschild spacetime (S-GUP). In particular, motion of test particles in innermost stable circular orbit (ISCO) around black hole in S-GUP is studied. Additionally, we have shown the mimic of Kerr and S-GUP BHs parameters. Finally, we have studied Quasi-periodic oscillation  around such a BH. By using observational data of few quasars, we have achieved constraints on the free parameter $\epsilon$ in the S-GUP spacetime. We have chosen $c=G=\hbar=1$ throughout our manuscript.

\section{Particle Dynamics in GUP-Modified Schwarzschild spacetime} 
\label{null}

\begin{figure}
    \centering
    \includegraphics[scale=0.6]{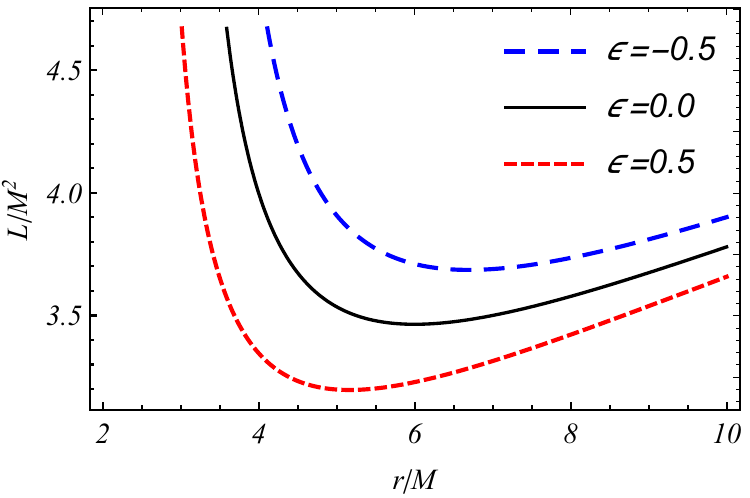}
    \includegraphics[scale=0.6]{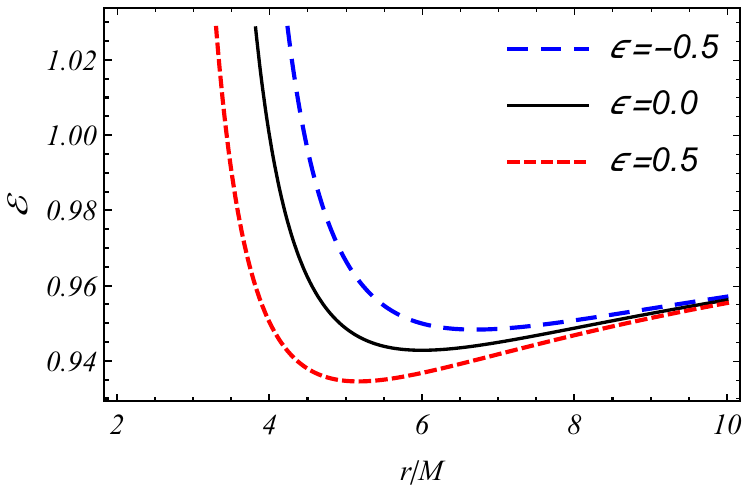}
    \caption{The graphs illustrate how the specific angular momentum and specific energy of test particles in their stable orbits (minimum energy) around a static BH in S-GUP gravity are analyzed. The graph suggests that the stable orbit adjusts slightly with variations in the parameter $\epsilon$.}
    \label{fig:enter-label}
\end{figure}

In this section, we investigate the movement of test particles near a BH with GUP correction.
In Boyer-Lindquist coordinates, the S-GUP metric is given by \cite{Scardigli2015EPJC,2024PDU....4301392H},
\begin{align}
    ds^2=-f(r) d t^2+f(r)^{-1} d r^2+r^2(d \theta^2 +\sin^2 \theta d\phi^2 ),
    \label{metric}
 \end{align}
with 
\begin{align}
   f(r) = 1-\frac{2 M}{r}+\epsilon\frac{M^2}{r^2},
\end{align}
where $\epsilon$ is characteristic parameter of GUP correction. We like to point out that the above metric is not derived formally by solving the governing field equations. Rather, it is proposed by previous authors to lowest order in $\epsilon$ as a test metric to understand the GUP effects in Schwarzschild spacetime \cite{Scardigli2015EPJC}.
The behavior of these test particles is described by the following Lagrangian:
 \begin{align}
    L_p=\frac{1}{2}mg_{\mu\nu}\Dot{x}^\mu\dot{x}^\nu\,
 \end{align}
where $m$ represents the particle's mass, and dot denotes differentiation with respect to proper time $\tau$.
It is important to highlight that $x^\mu(\tau)$ represents the world-line of the particle, parameterized by the proper time $\tau$ while the particle's four-velocity is defined as $u^\mu=\frac{dx^\mu}{d\tau}$. In a spherically symmetric spacetime, there are two Killing vectors associated with time-translation and rotational invariance of spacetime i.e., $\xi^\mu=(1,0,0,0)$ and $\eta^\mu=(0,0,0,1),$  respectively representing the total energy $E$ and angular momentum $L$ of the test particle. It can be formulated by:
\begin{align}
\mathcal{E}&=u_\mu\xi^\mu=-f(r)\Dot{t}, \nonumber\\
\mathcal{L}&=u_\mu\eta^\mu=r^2\sin^2\theta\Dot{\phi}.
\label{energy}
\end{align}
In Eqs.~(\ref{energy}) $\mathcal{E}=E/m $  and  $\mathcal{L}=L/m $ stand for the specific energy and specific angular momentum, respectively. The equation of motion for the test particle can be obtained by using normalization condition:
\begin{align}\label{normal}
    g_{\mu\nu}u^\mu u^\nu=\delta,
\end{align}
where $\delta=0$ and $\delta=\pm1$ lead to geodesics for massless and massive particles, respectively. Here $\delta=+1$ corresponds to spacelike while $\delta=-1$ is relevant for timelike trajectory. For massive particles, the equation of motion is governed by the timelike geodesics of spacetime, and these equations can be derived by utilizing Eq.~(\ref{normal}). Considering Eqs.~(\ref{energy}) and (\ref{normal}), one can derive the equation of motion in the following form:
\begin{align}
    \dot{r}^2=\mathcal{E}-f(r)\Big(1+\frac{{\mathcal{L}^2}}{r^2\sin^2{\theta} }\Big).
\end{align}
In a static and spherically symmetric spacetime, if a particle initially starts its motion in the equatorial plane than its motion remains confined in the equatorial plane throughout its motion. Thus restricting the motion of the particle to the plane, in which $\theta=\pi/2$ and $\Dot{\theta}=0$, the equation of the radial motion can be expressed in the form:
 \begin{align}
     \Dot{r}^2=\mathcal{E}^2-V_\text{eff},
 \end{align}
where the effective potential of the radial motion reads
\begin{align}
    V_\text{eff}=f(r)\Big(1+\frac{\mathcal{L}^2}{r^2\sin^2{\theta}}\Big).
\end{align}
Now, applying standard conditions for circular motion, 
$\dot{r}=0$, and $\ddot r=0$, we can derive expressions for the specific angular momentum and the specific energy for circular orbits at the equatorial plane ($\theta=\pi/2$) in the following form:
\begin{align}
    \mathcal{E}=\sqrt{\frac{(\epsilon M^2+(r-2M)r)^2}{r^2(r^2-3M r+2\epsilon M^2)}},
\end{align}
\begin{align}
    \mathcal{L}=\frac{r^2(M r-\epsilon M^2)}{r^2-3M r+2\epsilon M^2}.
\end{align}
Fig.~(\ref{fig:enter-label}) depicts the radial variations of specific energy and angular momentum of particle in circular orbits. One can easily see that specific energy and angular momentum  decrease when $\epsilon$ increases. Additionally, the  minimum values of both specific energy and angular momentum which corresponds to the location of Innermost Stable Circular Orbit (ISCO) decrease when $\epsilon$ increases. It is noteworthy that the behaviour of both conserved quantities is similar to the Schwarzschild case when $\epsilon=0$. 

If we reconsider expression (\ref{normal}) for the case $\delta=0$ and after some straightforward calculations following \cite{2021PhRvD.103j4070R}, we can find radius of the photon sphere as
\begin{align}     
r_\gamma=\frac{3M+\sqrt{9M^2-4 \epsilon M^2}}{2}.
\end{align}

\subsection{ Innermost stable circular orbits}

\begin{figure}
\includegraphics[scale=0.665]{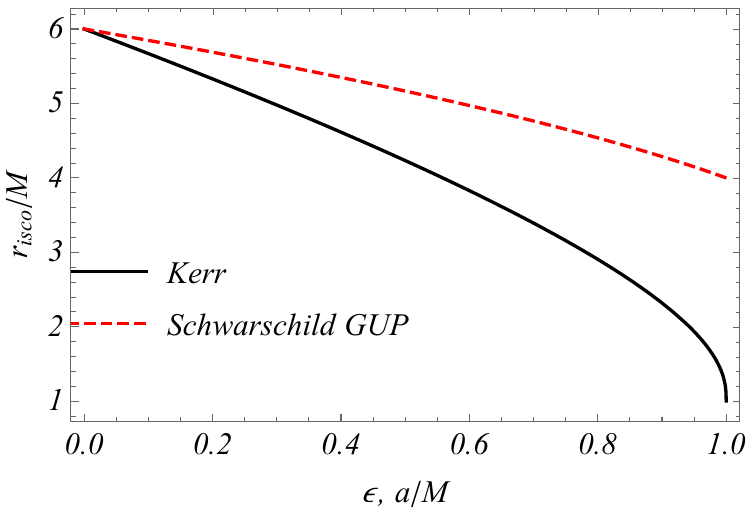}
\includegraphics[scale=0.45]{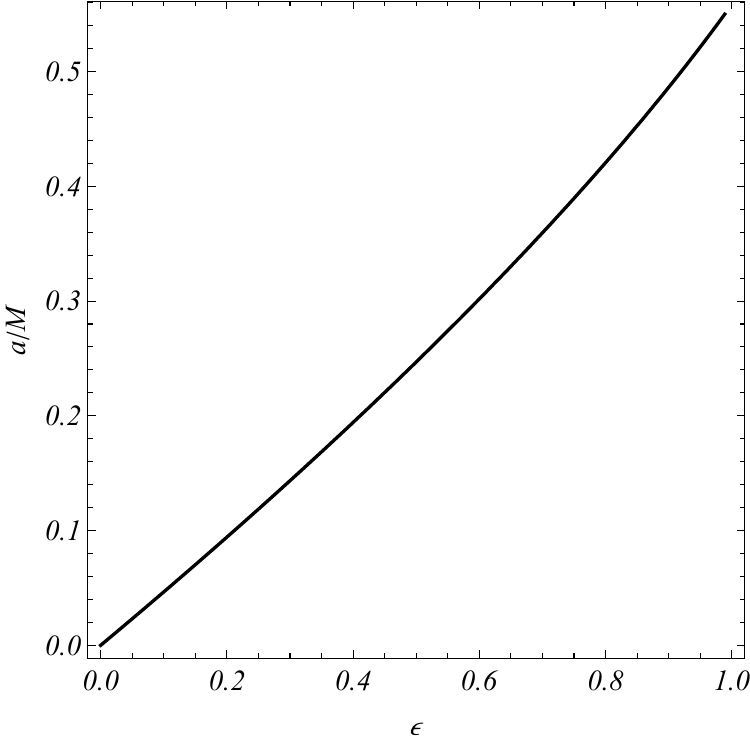}
    \caption{These graphs depict the relationship between ISCO positions, spin parameter $a/M$ and $\epsilon$, respectively(left panel) and the degeneracy between Kerr and S-GUP metric parameters which provide the same ISCO radius.}\label{risco}
\end{figure}

As mentioned earlier that ISCO is located at the minimum of the specific energy or angular momentum. This condition can be expressed as $\mathcal{E}_{,r}=0 \,\, \rm{or} \,\, \mathcal{L}_{,r}=0$ or in the explicit form as:
\begin{align}\label{iscoeq}
\frac{\sqrt{M} (-6 M r^2 + r^3 + 9 M^2 r \epsilon - 4 M^3 \epsilon^2)}{2 \sqrt{r - M \epsilon} \left( -3 M r + r^2 + 2 M^2 \epsilon \right)^{3/2}}=0.
\end{align}
By solving Eq.~\ref{iscoeq} with respect to $r$, the position of the ISCO radius of test particles is obtained in the following form:
\begin{align}
   r_{ISCO}=2M+\frac{M^2(3\epsilon-4)}{Z^\frac{1}{3}}-Z^\frac{1}{3};\  \  \\
   Z=M^3(-8+9\epsilon-2\epsilon^2)+\sqrt{M^6\epsilon^2(5-9\epsilon+4\epsilon^2)}.\nonumber
\end{align}
In order to analyse the behaviour of ISCO radius, we illustrate dependency of ISCO radius $r_{ISCO}$ on $\epsilon$. Fig.~(\ref{risco}), displays the dependency of ISCO radius on the spin $a/M$ (for Kerr BH) and the GUP parameter $\epsilon$ (for the S-GUP BH), respectively. From the left panel of these pictures it follows that the spin parameter $a/M$ of rotating Kerr BH mimics the $\epsilon$ of S-GUP BH. The effect of $\epsilon$ parameter can be figured out as a Kerr BH with smaller spin coefficient.
\section{Fundamental frequencies around S-GUP spacetime}
In this section, we derive the fundamental frequencies governing the motion of a particle orbiting around a BH with additional structures such as GUP. Specifically, we analyze the frequencies related to Keplerian orbits, as well as the radial and vertical oscillations of particle's orbits. These frequencies are essential for investigating models of QPOs.

\subsection{Keplerian frequencies}
The angular velocity of the particles around the BH, as measured by an observer located at infinity, is referred as the orbital (Keplerian) frequency 
$\Omega_\phi$, defined as 
$\Omega_\phi=\frac{d\phi}{dt}$. By using this definition, we can get following equation for orbital frequency in spherically symmetric spacetime~\cite{2024PDU....4601569D},
\begin{align}
    \Omega_\phi=\sqrt{\frac{-\partial_r g_{tt}}{\partial g_{\phi\phi}}}=\sqrt{\frac{f'(r)}{2r}}.
\end{align}
Specifically, if we use metric (\ref{metric}), the last expression takes the following form: 
\begin{align}
\Omega_\phi=\frac{\sqrt{M(r-M\epsilon)}}{r^2}.
\end{align}
Furthermore, to express the frequencies in Hertz (Hz) we use the following equation:
\begin{align}
    \nu_\phi=\frac{c^3}{ 2\pi G M}\frac{\sqrt{M(r-M\epsilon)}}{r^2}.
\end{align}
\begin{figure}
\centering
\includegraphics[scale=0.6]{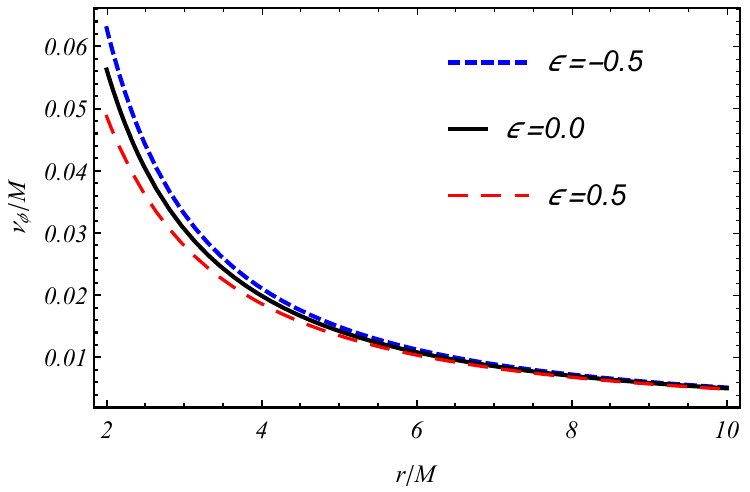}
\includegraphics[scale=0.6]{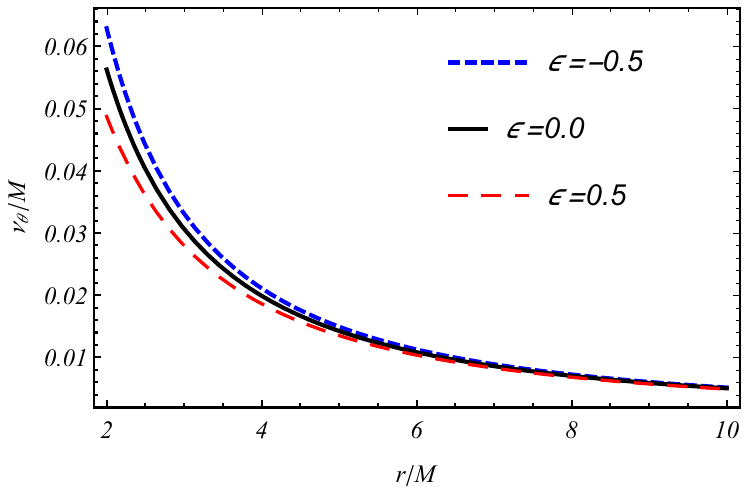}
\includegraphics[scale=0.6]{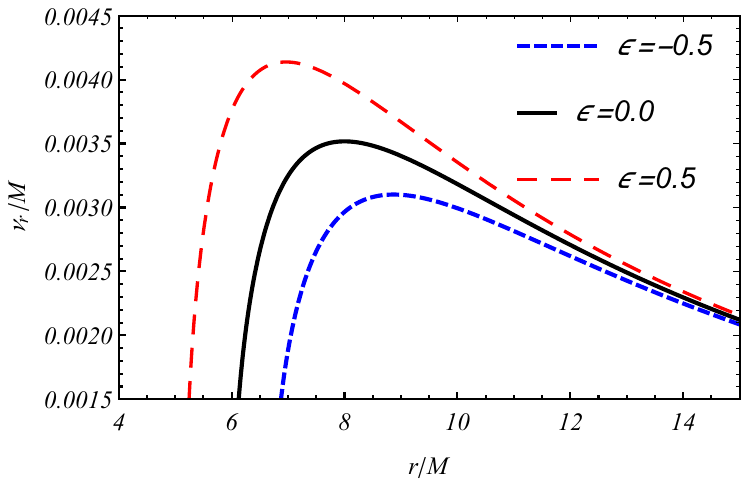}
\caption{The figure displays the radial dependence of Keplerian, vertical and radial frequencies for test particles around a S-GUP BH for different values of parameter $\epsilon$.}
    \label{frequencies}
\end{figure}
\subsection{Harmonic oscillations}
We analyze a test particle oscillating along the radial, angular, and vertical axes within its stable orbit around a static BH situated in the equatorial plane, resulting from minor deviations from these orbits as $r_0+ \delta r$ and $\theta_0+\delta\theta$, where $\theta_0=\pi/2$. The frequencies of radial and vertical oscillations, observable by a distant observer, can be determined using harmonic oscillator equations as \cite{Bambi2017book}:
\begin{align}
    \frac{d^2\delta r}{dt^2}+\Omega^2_r\delta r=0,\  \frac{d^2\delta\theta }{dt^2}+\Omega^2_\theta \delta\theta=0,
\end{align}
where
\begin{align}
 \Omega_r^2=-\frac{1}{2g_{rr}\Dot{t}^2}\partial_r^2 V_\text{eff}(r,\theta)\Big\arrowvert_{\theta=\pi/2},
\end{align}
\begin{align}
        \Omega_\theta^2=-\frac{1}{2g_{\theta\theta}\Dot{t}^2}\partial_\theta^2 V_\text{eff}(r,\theta)\Big\arrowvert_{\theta=\pi/2},
\end{align}
are the frequencies of the radial and vertical oscillations, respectively. By using above equations, we derive expressions for the frequencies within the spacetime of a static BH as follows:
\begin{figure}
\centering
\includegraphics[scale=0.6]{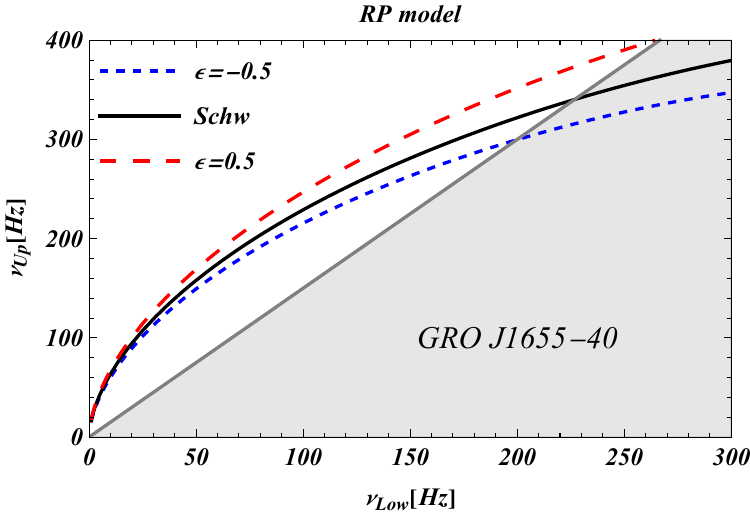}
\includegraphics[scale=0.6]{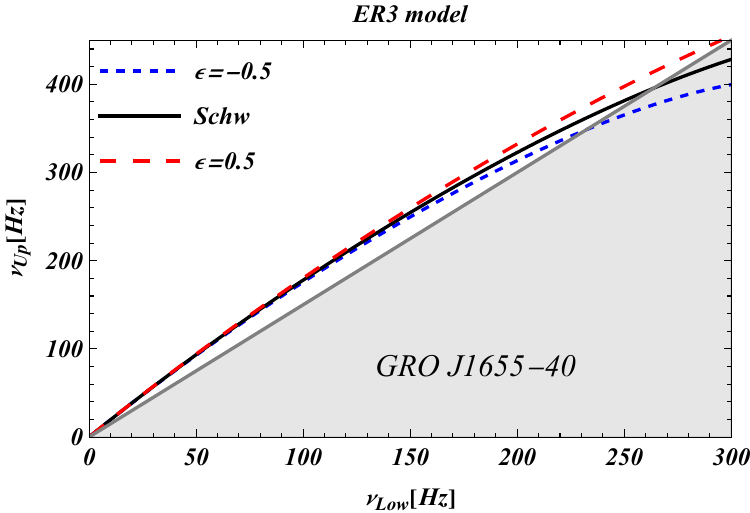}
\includegraphics[scale=0.6]{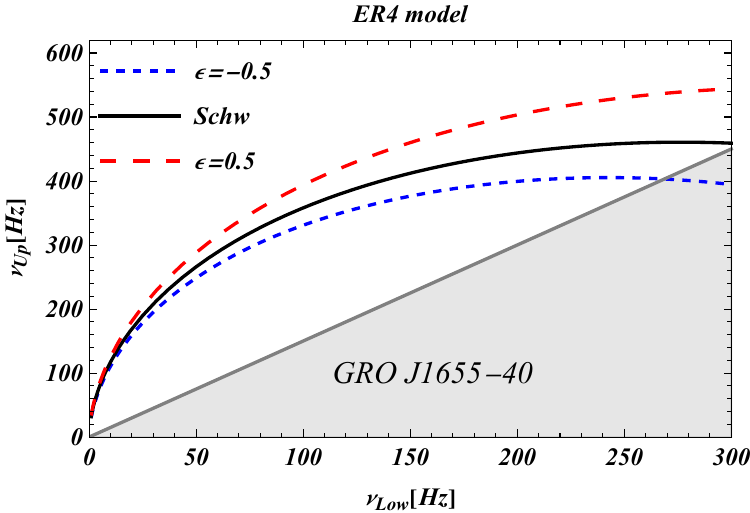}
\includegraphics[scale=0.6]{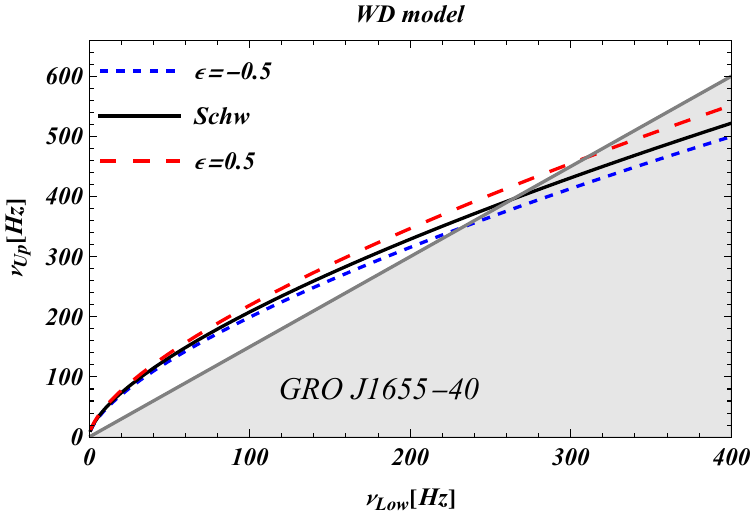}
         
\caption{The relationship between the upper and lower frequencies of twin peak QPOs in the spacetime around BHs in S-GUP in the RP, ER and WD  models for the different values of the parameter $\epsilon$.}
        \label{UpLow1}
    \end{figure}
    
    \begin{figure}
        \centering
        \includegraphics[scale=0.6]{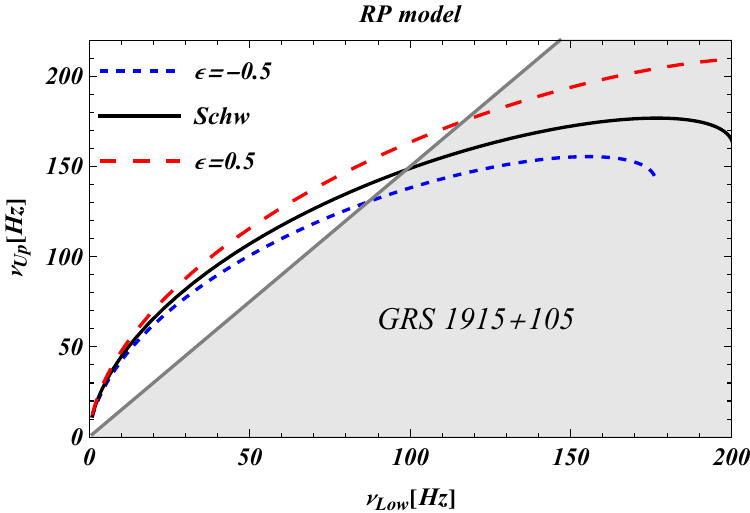}
         \includegraphics[scale=0.6]{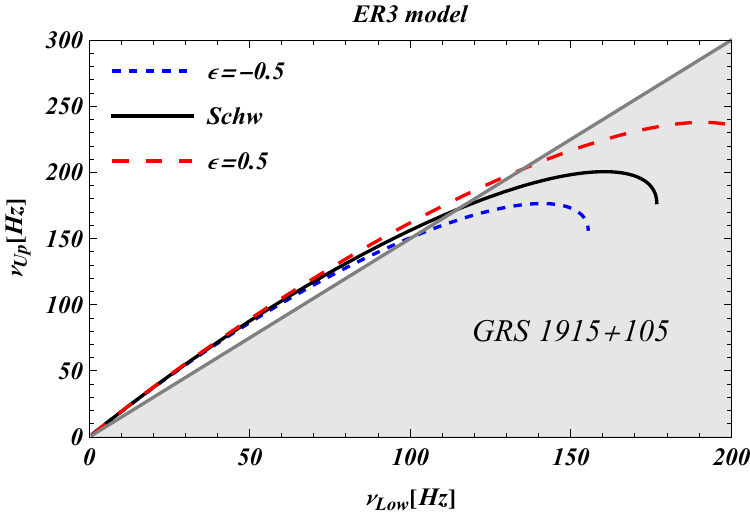}
         \includegraphics[scale=0.6]{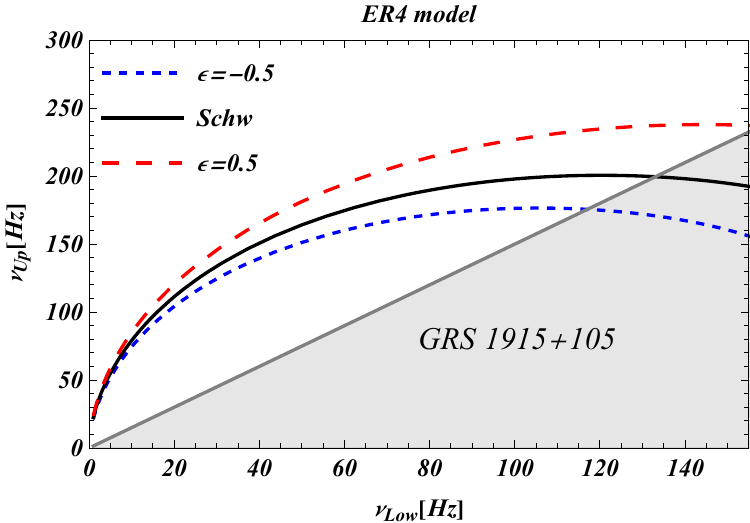}
         \includegraphics[scale=0.6]{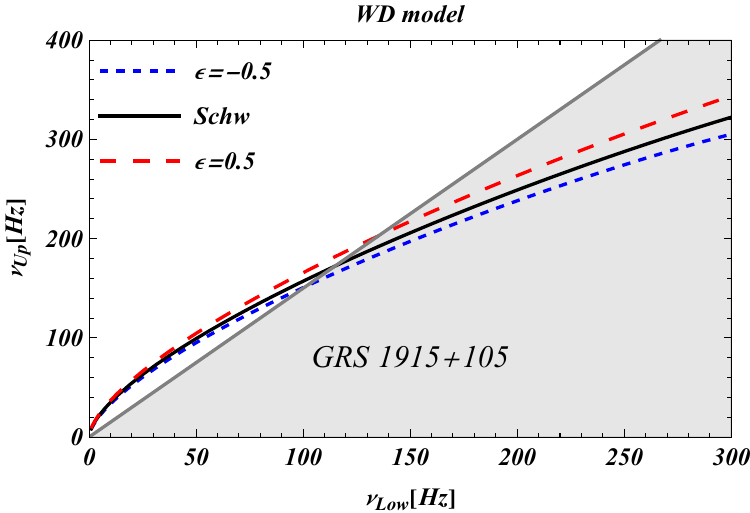}
         
        \caption{The same figure as Fig.~(\ref{UpLow1}) in the case of GRS 1915+105.}
        \label{UpLow2}
    \end{figure}
     \begin{figure}
        \centering
        \includegraphics[scale=0.6]{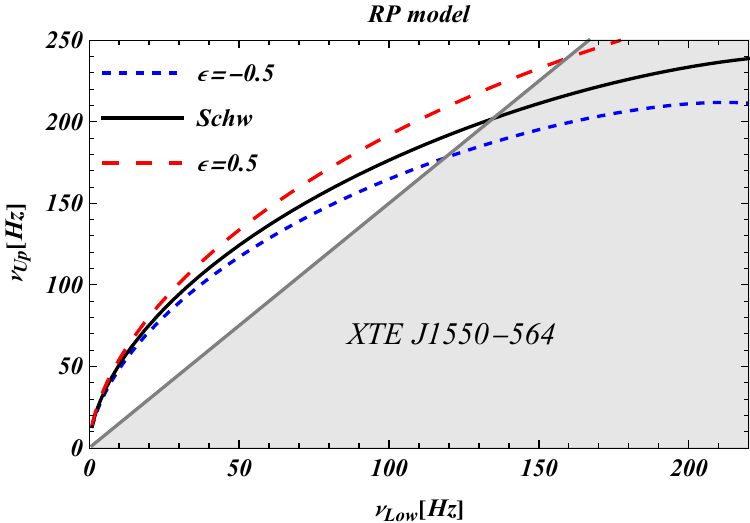}
         \includegraphics[scale=0.6]{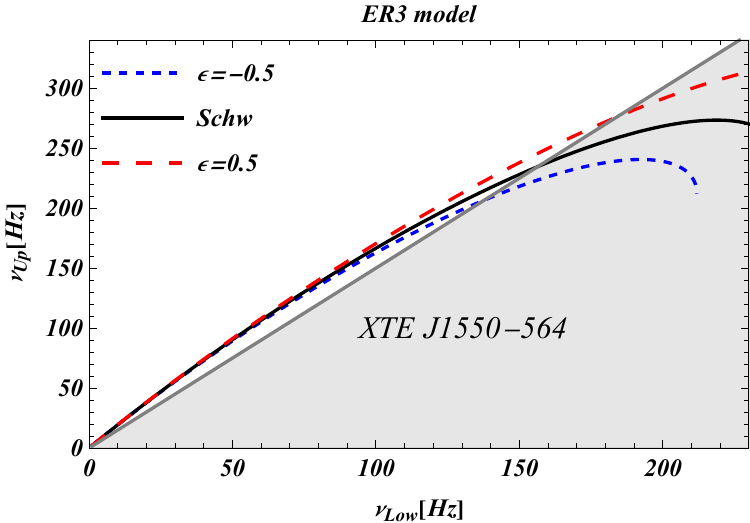}
         \includegraphics[scale=0.6]{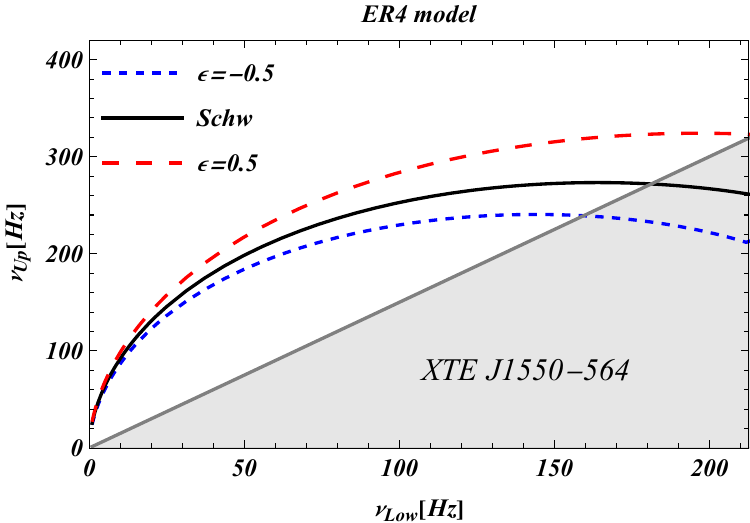}
         \includegraphics[scale=0.6]{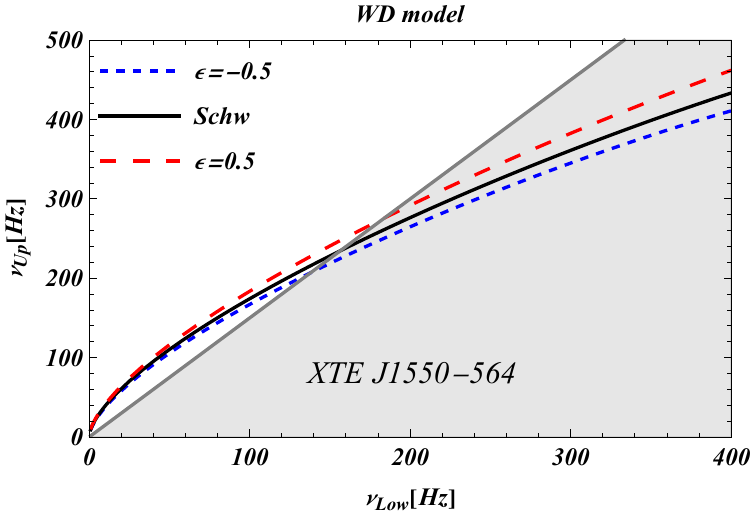}
         
        \caption{The same figure with Fig.~(\ref{UpLow1}) in the case of XTE J1550-564.}
        \label{UpLow3}
    \end{figure}
     \begin{figure}
        \centering
        \includegraphics[scale=0.6]{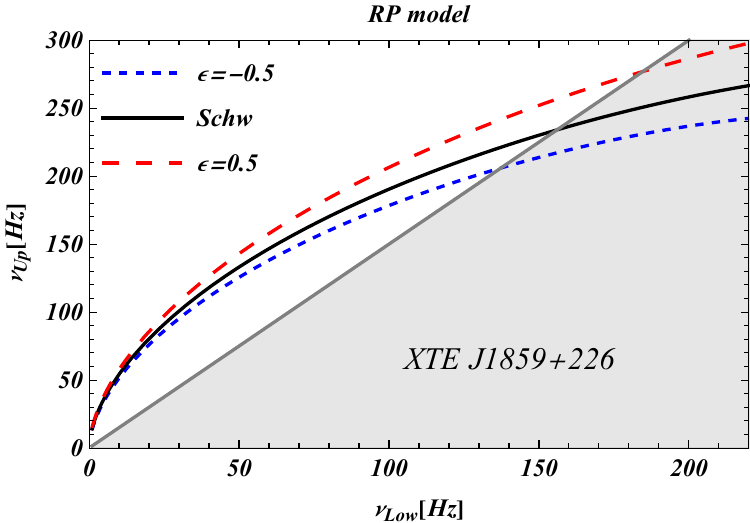}
         \includegraphics[scale=0.6]{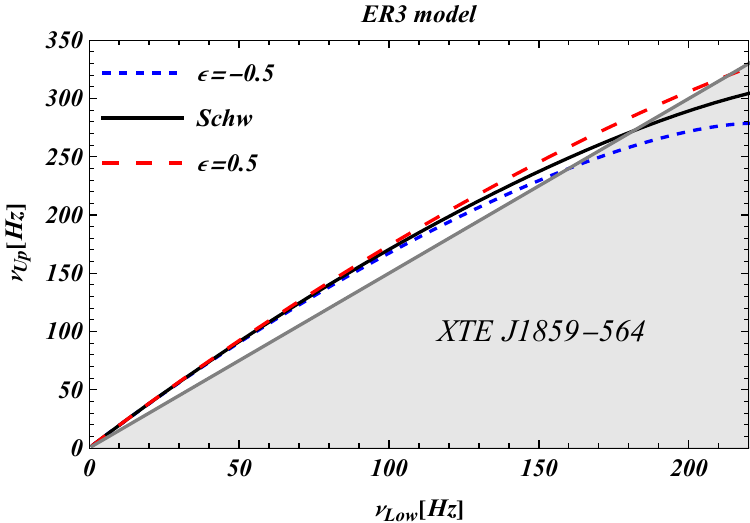}
         \includegraphics[scale=0.6]{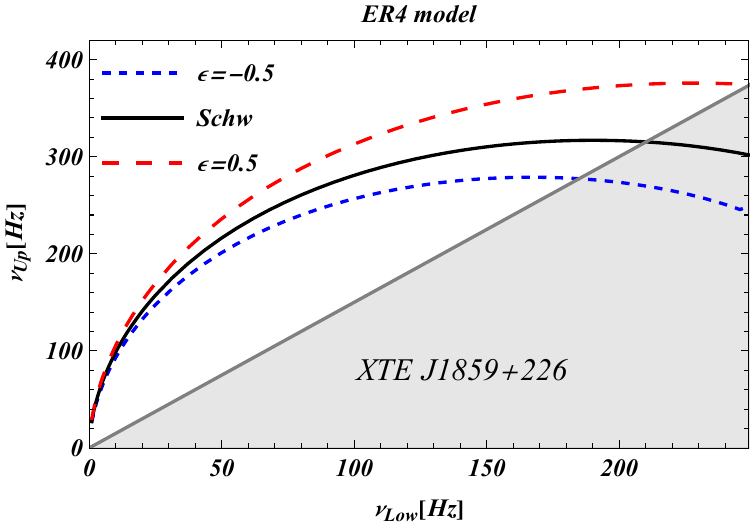}
         \includegraphics[scale=0.6]{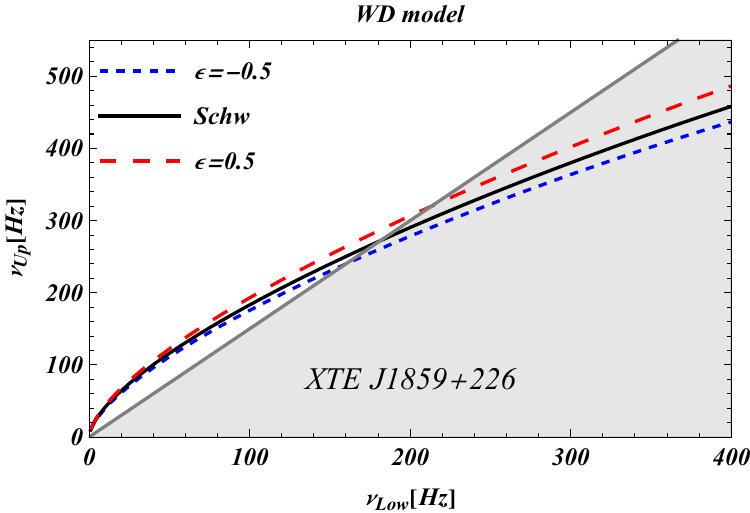}
         
        \caption{The same figure with Fig.~(\ref{UpLow1}) in the case of XTE J1859+226.}
        \label{UpLow4}
    \end{figure}
\begin{align}
    \nu_r&=\frac{c^3}{2\pi G M}\frac{\sqrt{M r^2(r-6M)-4\epsilon^2 M^2+9\epsilon M^3 r}}{r^3}, \nonumber \\
    \nu_\theta&=\nu_\phi.
\end{align}
From Fig.~(\ref{frequencies}) we find  that the GUP correction affects negatively for both orbital and vertical frequencies. Conversely, the GUP effect is positive for the case of radial frequency as well as one can observe that maximum value of the radial frequency increases with increasing parameter $\epsilon$.

\section{Astrophysical applications}
This section is dedicated to investigate the potential frequency values of twin-peak QPOs around a Schwarzschild geometry modified by the GUP. This exploration involves employing QPO models to analyze and characterize the observed frequencies. Furthermore, we also focus on determining the relationship between the mass of the BH and the parameter $\epsilon$ using their observational frequency data from QPOs. Additionally, we take into account BHs situated at the center of microquasars, namely GRO J1655-40, GRS 1915+105, XTE 1550-564, and XTE 1859+226.
\subsection{QPO models}
\begin{figure}
        \centering
        \includegraphics[scale=0.6]{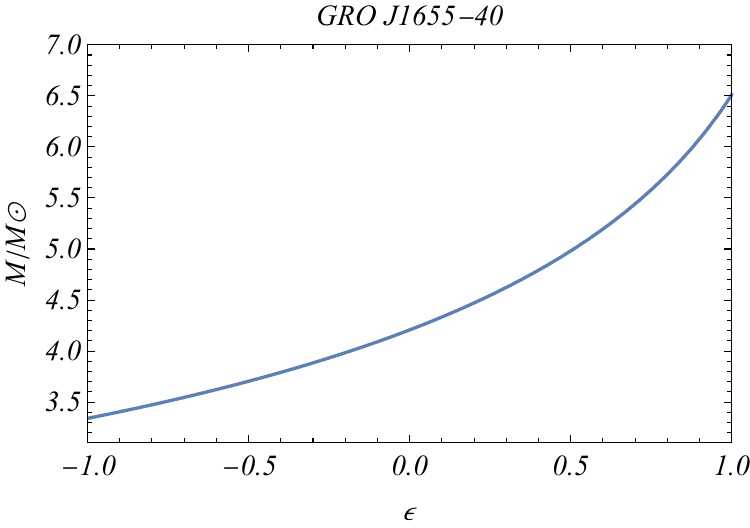}
         \includegraphics[scale=0.6]{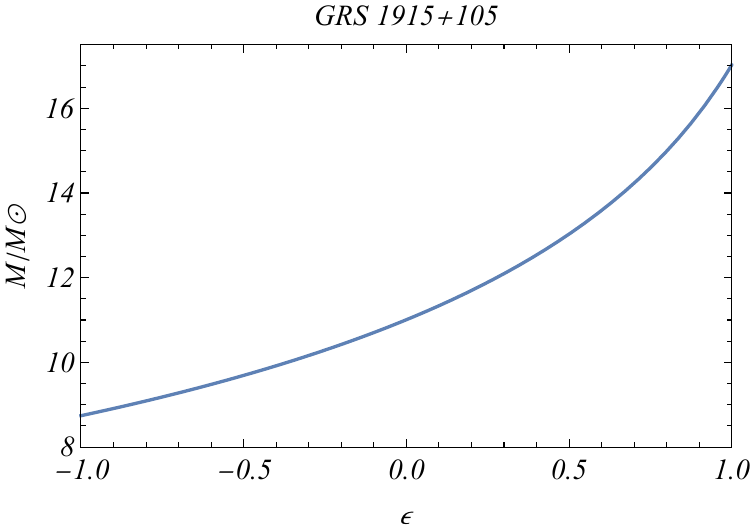}
          \includegraphics[scale=0.6]{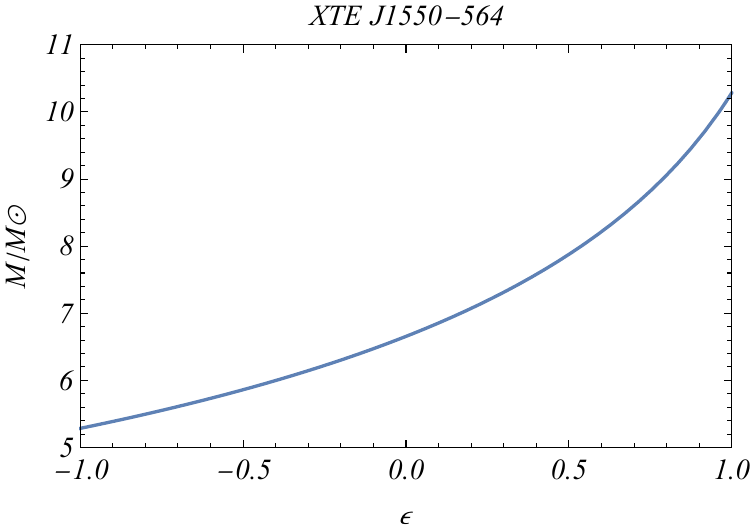}  
          \includegraphics[scale=0.6]{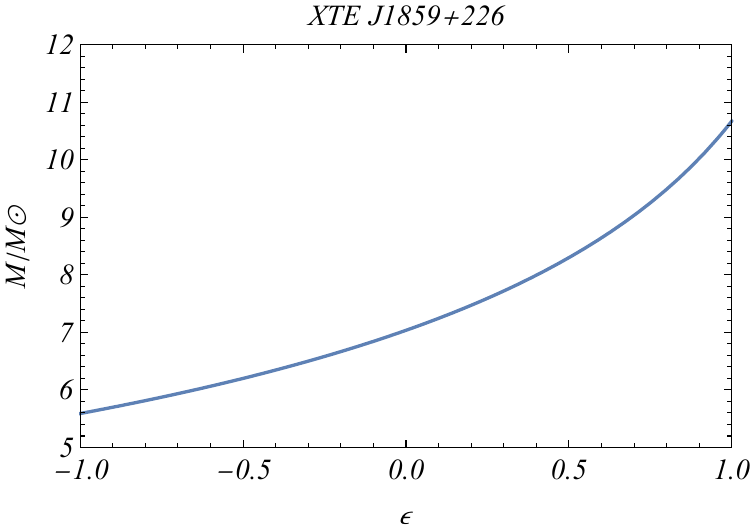}
        \caption{Estimated Mass of the centeral BHs in the heart of microquasars through QPOs in RP models as a function of GUP parameter $\epsilon$}
        \label{Mass cons}
    \end{figure}
In this subsection, our objective is to examine possible values for the upper and lower frequencies using models that describe twin-peak High-Frequency QPOs (HF QPOs). These models are based on the fundamental frequencies of particles in orbit around compact gravitating objects \cite{Shaymatov:2023jfa,Shaymatov:2022enf,Rayimbaev:2022mrk,Jusufi:2020odz}. We consider following models:
\begin{itemize}
\item The Relativistic Precession (RP) model, initially proposed by Stella and Vietri, was introduced to explain kHz twin-peak QPOs with frequencies ranging from 0.2 to 1.25 kHz observed in neutron stars within Low-Mass X-ray Binary systems (LMXRBs)\cite{1998tx19.confE.315S}. Subsequently, it was demonstrated that this model is applicable to BH candidates in binary systems involving both BHs and neutron stars. Ingram and Motta \cite{A.Ingram10.1093/mnras/stu15850} later expanded the RP model, aiming to achieve precise measurements of the mass and spin of central BHs in microquasars, utilizing data from the power density spectrum of the BH accretion disk.
According to the RP model, the upper and lower frequencies are characterized by the frequencies of radial, vertical, and orbital oscillations, represented as $\nu_U=\nu_\phi$ and $\nu_L=\nu_\phi-\nu_r$, respectively.
\item The Epicyclic Resonance (ER) model explores resonances arising from axisymmetric oscillation modes of a thin accretion disc around BHs. In this model, the frequencies of the disc oscillation modes are intricately connected to the frequencies of the orbital and epicyclic oscillations associated with the circular geodesics of test particles.
Here, we consider variations of the ER model, namely, ER3, and ER4, each distinguished by their specific oscillation modes. The upper and lower frequencies in ER3-4 modes are defined as follows \cite{2021EPJC...81..699R}:
ER3 Mode:
$\nu_U = \nu_\theta + \nu_r$,
$\nu_L = \nu_\theta$
ER4 Mode:
$\nu_U = \nu_\theta + \nu_r$,
$\nu_L = \nu_\theta - \nu_r$
These frequency expressions characterize the behavior of the oscillation modes in the ER3, and ER4 variations of the model \cite{2002A&A...396L..31A}.
\item The Warped Disc (WD) model is built on the assumption of non-axisymmetric oscillatory modes within a thin accretion disc surrounding BHs and neutron stars \cite{Kato2007ResonantEO,Kato2008FrequencyCO}. In the WD model, the upper and lower frequencies are defined as follows:
$\nu_U = 2\nu_\phi - \nu_r$,
$\nu_L = 2(\nu_\phi - \nu_r)$ \cite{Kato2007ResonantEO,Kato2008FrequencyCO}.
Here, the vertical oscillatory frequency $\nu_\theta$ is introduced through the assumptions of vertical axially symmetric oscillations in the accretion disc, leading to disc warping. This model considers the impact of non-axisymmetric modes on the oscillatory behavior of the accretion disc around compact gravitating objects.
\end{itemize} 
In Figs.~(\ref{UpLow1})-(\ref{UpLow4}), we illustrate the $\nu_U-\nu_L$ diagram for twin-peak QPOs around S-GUP BH in the Relativistic Precession (RP), Epicyclic Resonance (ER3-ER4), Warped Disc (WD) models. To get these pictures we used optical measurements data for BHs mass in the center of microquasars GRO J1655-40 (Fig.~\ref{UpLow1}), GRS 1915+105 (Fig.~\ref{UpLow2}), XTE J1550-564 (Fig.~\ref{UpLow3}) and XTE J1859+226 (Fig.~\ref{UpLow4}), respectively. It is apparent that upper frequencies in the presence of the GUP parameter can be higher than the schwarzschild BH case. Through analysing these figures, we can conclude as follows: ER4 model of QPOs could explain high-frequency QPOs both in the schwarzschild (up to more than 400 Hz for GRO J1655-40) and in S-GUP with positive values of $\epsilon$ (up to more than 500 Hz for GRO J1655-40). Additionally, we can see that RP, ER3 and WD models could explain QPOs with frequency range of approximately from 150 Hz (for GRS 1915+105) to 400 Hz (for GRO  J1655-40). One should notice that distinguishing BH with GUP modification from Schwarzschild BH depends on model. In Figs. (\ref{UpLow1}-\ref{UpLow4}), the area between black and red dotted lines implies the possible values of frequency of the twin-peak QPOs from S-GUP BHs, while the areas between both black and blue dotted or blue dotted and grey lines may correspond S-GUP or Schwarzschild BHS. The region under grey lines corresponds to twin-peak QPOs which cannot be observed.

    \begin{figure}
        \centering
        \includegraphics[scale=0.6]{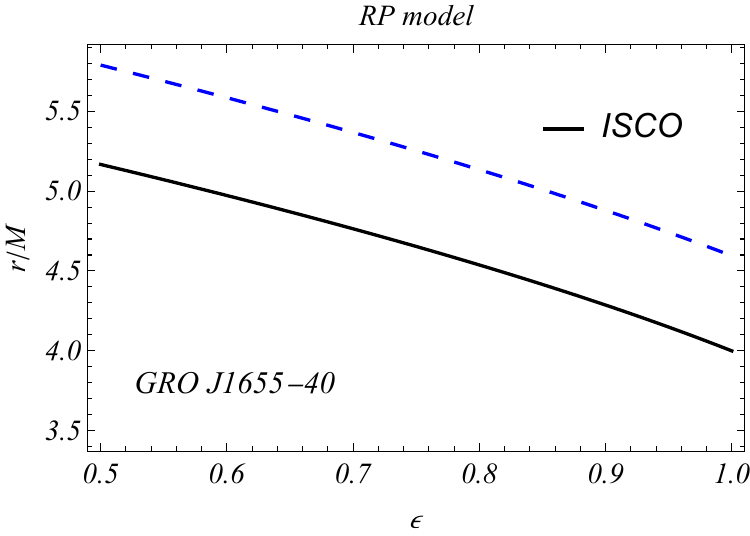}
        \includegraphics[scale=0.6]{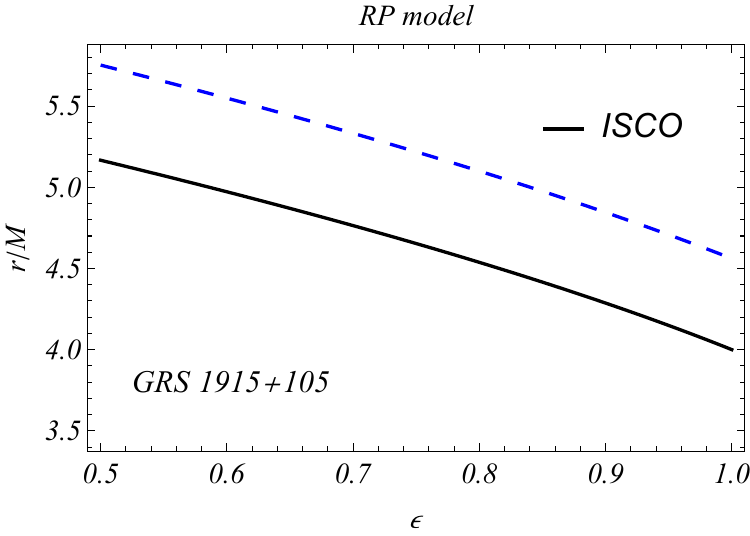}
        \includegraphics[scale=0.6]{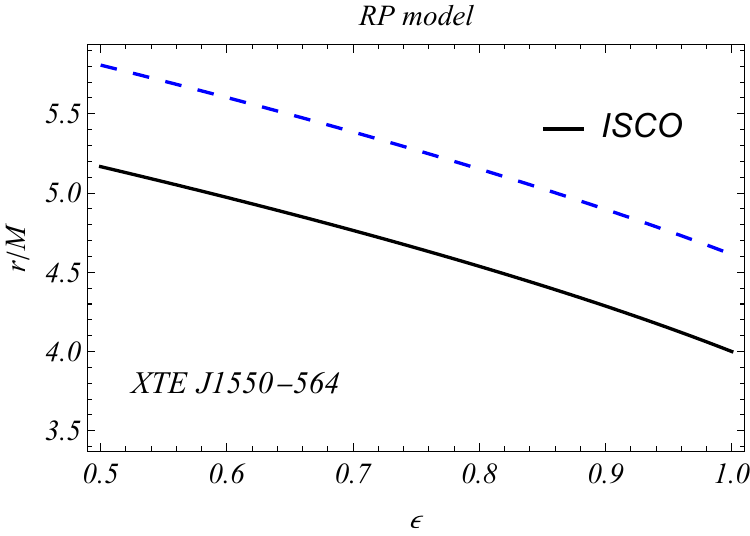}
        \includegraphics[scale=0.6]{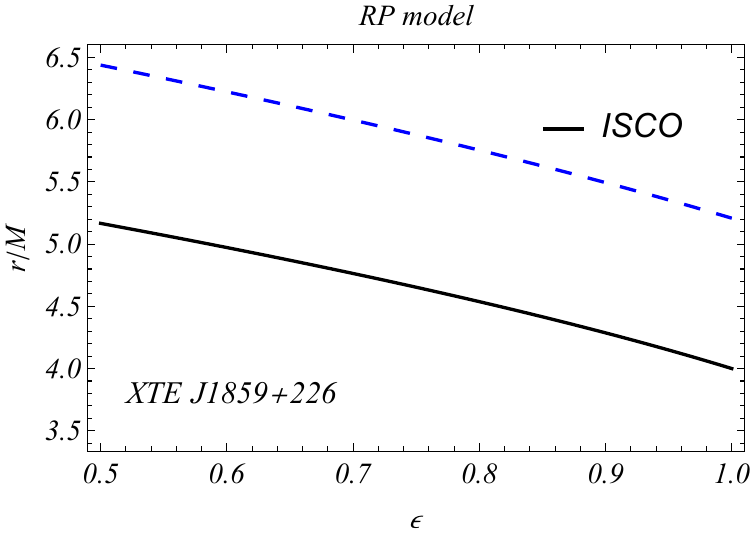}
        \caption{Dependency of location of radiation on parameter $\epsilon$ under RP models for different BHs at the center of different microquasars.}
        \label{rad}
    \end{figure}
    
\subsection{BH mass constraints using QPO frequencies}

In this section, we obtain constraints on the mass and the parameter $\epsilon$ of the S-GUP BH at the center of the microquasars GRO J1655-40, GRS1915+105, XTE J1550-564 and XTE J1859-226 graphically. To establish the relationship between the BH mass and the parameters $\epsilon$, we formulate equations for the upper and lower frequencies utilizing the QPO radius, which can be numerically determined as follows:
\begin{align}
    \nu_L(r,q,M)=\nu_L^{ob},\  
    \nu_U(r,q,M)=\nu_U^{ob}\,,
\end{align}
where $\nu_L^{0b}$ and $\nu_U^{0b}$ are observational data of the lower and upper frequencies \cite{2023AnPhy.45469335R}. 
Fig.~(\ref{Mass cons}) shows the possible relationship between BH's mass and $\epsilon$ for different BH candidates. One can easily see that the prediction for the BH mass at the center of microquasar GRO J1655-40, can get smaller values. Conversely, the higher mass prediction is valid for the case of GRS 1915-105. Generally, for all BH candidates we have considered the mass prediction get higher value with $\epsilon$ increase. In Fig.~(\ref{rad}) we illustrate the relationship between the ISCO radius and $\epsilon$ in the RP model for four different QPO sources. It is evident from this figure that as $\epsilon$ decreases, the radius increases. Unfortunately, since we have only two observable QPOs, exact amount of mass, radius and $\epsilon$ can not be found at the same time. However, we can use results of another measurements for the mass of the BH's candidates in order to set exact constraint for $\epsilon$ and radius. In this way we have composed Table~(\ref{Table}), where we have provided mass of the sources from another measurements, and have got constraint for GUP parameter $\epsilon$. From this table, we can see that the GUP parameter $\epsilon$ can get higher value in the case of XTE J1550-564 while it gets the smallest value in the case of XTE J1859+226. In \cite{Martin}, authors show that if we assume the BH in the center of the microquasar GRO J1655-40 is Kerr one, by using RP model, the best-fitted value of $r/M$ and $a/M$ are $5.68\pm0.05$ and $0.286\pm0.004$, respectively. By comparing this result, one can be observed that $\epsilon$ parameter can mimic Kerr BH with smaller spin coefficient. Also, the S-GUP model predicts that the origin of the high frequency QPO is closer to the BH in this model than in the Kerr model.


\begin{table}[h]
   \centering
\begin{tabular}{|c|c |c |c |c|c|c|}
    \hline
    Sources & $M/M\odot $& $r/M$ &$\epsilon$ \\
    \hline
    GRO J1655-40 & 5.4±0.3 \cite{Motta2014MN} & 5.42781±0.24 &0.684221$^{+0.12}_{-0.10}$ \\\hline
    XTE J1550-564 & 9.1± 0.61 \cite{Remillard2002APS}& $5.17777^{+0.28101}_{-0.3165}$& 0.808439 $ ^{+0.136062}_{-0.108749}$ \\
    \hline
    XTE J1859+226 & 7.85±0.46 \cite{M10.1093/mnras/stac2142} &  6.75744 $^{+0.183572}_{-0.154463}$   &0.348667$^{+0.183572}_{-0.154463}$\\
    \hline
    GRS 1915+105 & 12.4 \cite{25article} & 6.07799$_{-0.85756}^{+0.72}$   &0.37048$^{-0353219}_{+0.50621}$ \\
    \hline
     H1743-322&  $\geq$ 9.29 \cite{A.Ingram10.1093/mnras/stu1585}& $\leq$ 5.75301 &  $\geq$0.505187 \\
    \hline
\end{tabular}\caption{The table includes names of source and corresponding parameters. Here we used the photometric mass measurements' results for masses of the BHs in the center of the quasars(second column). Third and last column represents best fitted values of $r/M$ and parameter $\epsilon$.}\label{Table}
\end{table}

\section{Conclusion}\label{conlusion}
 This work primarily focuses on comparing the gravitational effects on test particles motions around the Schwarzschild and S-GUP BHs.
 \begin{itemize}
 \item We discussed the radial variations of specific energy and angular momentum within circular orbits which reveal a noticeable trend: both specific energy and angular momentum decrease as $\epsilon$ increases. Furthermore, the minimum positions of both specific energy and angular momentum, corresponding to the location of the ISCO, decrease with increasing $\epsilon$. Importantly, it is worth noticing that the behavior of both quantities revert to  Schwarzschild case when $\epsilon=0$.
\item  We further discussed the spin parameter $a/M$ of rotating Kerr BH mimics the $\epsilon$ of S-GUP BH. The effect of $\epsilon$ parameter can be figured out as a Kerr BH with smaller spin coefficient. In addition we have studied degeneracy between spin parameter in Kerr and $\epsilon$ parameter in S-GUP spacetimes. 
 \item From Fig.~(\ref{frequencies}), we can conclude that the GUP correction has a negative impact on both orbital and vertical frequencies. Conversely, its effect is positive for radial frequency.
\item  It is evident that the prediction of mass for the BH at the center of the microquasar, known as GRO J1655-40, can assume smaller values. Conversely, the highest prediction for mass is applicable to the case of GRS 1915-105. Generally, for all BH candidates considered, the mass increases with an increase in $\epsilon$. We have compiled Table~(\ref{Table}). This table provides the masses of the sources from other measurements, and we have obtained constraints for the GUP parameter, $\epsilon$. From this table, it can be observed that the GUP parameter $\epsilon$ can assume higher values in the case of XTE J1550-564.
\end{itemize}
     
\section*{Acknowledgements} This research is partly  
supported by Research Grant F-FA-2021-510 of the Uzbekistan Ministry for Innovative Development.

\bibliographystyle{apsrev4-1}  
\bibliography{paper,gl,LE}

\end{document}